\documentclass[a4paper]{article}
\usepackage{graphicx}
\date{21 August 1998 }

\begin{document}

\title{Redshifts of New Galaxies}

\author{Halton ARP\\
  Max-Planck-Institut f\"ur Astrophysik\\ Karl-Schwarzschild-Str.~1\\
  85740 Garching, Germany}

\maketitle

\begin{abstract}
Observations increasingly demonstrate the spatial association of high
redshift objects with larger, low redshift galaxies. These companion
objects show a continuous range of physical properties - from very
compact, high redshift quasars, through smaller active galaxies and
finally to only slightly smaller companion galaxies of slightly higher
redshift. The shift in energy distribution from high to low makes it
clear that are seeing an empirical evolution from newly created to
older, more normal galaxies.

In order to account for the evolution of intrinsic redshift we must
conclude that matter is initially born with low mass particles whose
mass increase with time (age). This requires a physics which is
non-local (Machian) and which is therefore more applicable to the
cosmos than the Big Bang extrapolation of local
physics. Ambartsumian's "superfluid" foresaw some of the properties of
the new, low particle mass, protogalactic plasma which is required,
demonstrating again the age-old lesson that open minded observation is
much more powerful than theoretical assumptions.

Since the ejected plasma, which preferentially emerges along the minor
axis of the parent galaxy, develops into an entire galaxy, accretion
disks cannot supply sufficient material. New matter must be created
within a "white hole" rather than bouncing old matter off a "black hole".
\end{abstract}

\section{Introduction}

Evidence for the association of high redshift objects with low
redshift galaxies emerged in 1966 with completion of the systematic
study in the Atlas of Peculiar Galaxies. For the most recent summary
of the evidence to date see Arp (1998b). In the present report,
however, we concentrate on the most recent observations and those
which summarize best the empirical properties which characterize the
associations. These latest discoveries reinforce a picture of empirical 
evolution which progresses from newly born, high redshift protogalaxies 
(quasars) to older, more normal, low redshifts galaxies.

\section{The Quasars around NGC5985}

Fig. 1 shows one of the most exact alignments of quasars and galaxies
known. Attention was drawn to this region when it was discovered that
a very blue galaxy in the second Byurakan Survey (Markarian et
al. 1986) had a quasar of redshift z = .81 only 2.4 arcsec from its
nucleus (Reimers and Hagen 1998). Even multiplying by 3 x 104 galaxies
of this apparent magnitude or brighter in their survey they estimated
only a chance proximity of $10^{-3}$. (Nevertheless they took this as
proof that it was a chance projection! Also it was not referenced that
G.Burbidge, in 1996 in the same Journal, had published extensive list
of other quasars improbably close to low redshift galaxies).

\begin{figure}[ht]
\center{\includegraphics[width=1.0\hsize]{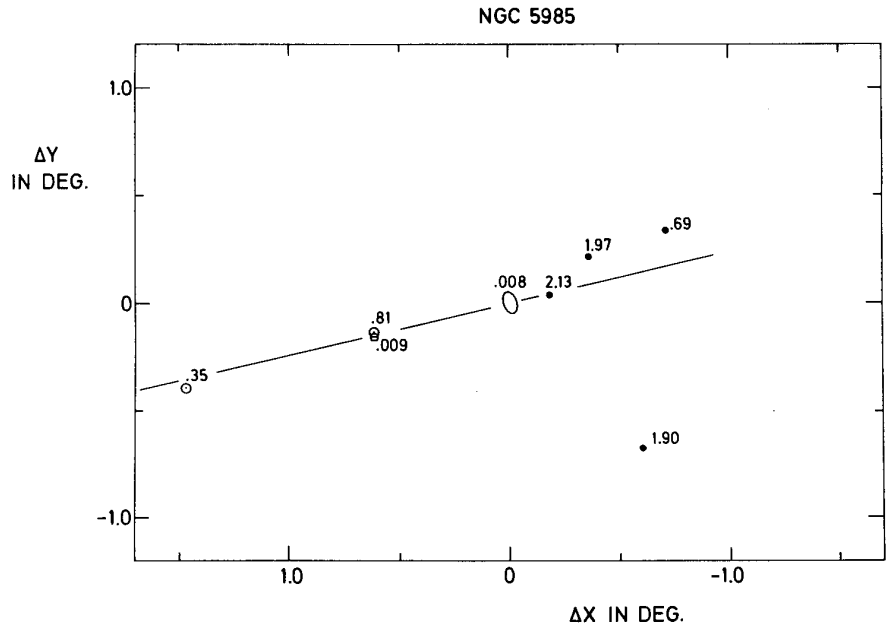}}
\caption{\footnotesize All catalogued active galaxies and quasars within 
the pictured area are plotted.Redshifts are labeled. The dwarf spiral
2.4 arcsec from the z = .81 quasar has z = .009 which marks it as
a companion of the Seyfert NGC5985 at z = .008. The line
represents the position of the minor axis of NGC5985.}
\end{figure}

But the galaxy was a dwarfish spiral, showing no active nucleus from
which the quasar could have emerged. Proceeding on the by now
overwhelming evidence that Seyfert galaxies eject quasars (Radecke
1997; Arp 1997a) I looked in the vicinity for a Seyfert. There was
NGC5985, only 36.9 arcmin away on the sky! The chance of finding a
Seyfert as bright as V = 11.0 mag. this close to the z = .81 quasar
was less than $10^{-3}$.

The next obvious question was: Were there other quasars in the field?
Fig. 1 shows that there are 6 catalogued quasars discovered in a
uniform search of the area in the Second Byurakan Survey. (P. Veron,
in this Symposium, reports 66 - 88 {\%} completeness for this survey -
probably typical for quasar surveys). But five of these six quasars
fall on a line through the Seyfert, with the dwarf galaxy along the
same line. What is the probability that three or four objects would
accidentally align within a degree or so of as straight line through
the Seyfert? Conservatively this can be computed to be of the order of
$10^{-4}$ to $10^{-5}$. But the most astonishing result of all is that
if one looks up the position of the minor axis in NGC5985 it turns
out, as shown in Figs. 1 and 2, to be a line that looks as though it
were drawn through the positions of the objects!

Just a simple visual evaluation of Fig. 1 would lead to the conclusion
that the configuration was physically significant. A combined
numerical probability of the configuration gives a chance of around
$10^{-9}$ to $10^{-10}$ of being accidental (Arp 1998). Nevertheless
several peer reviewers recommended against publication on the grounds
that the accidental probability was "greater" than this. But, of
course, several dozens of cases of anomalous associations had been
reported since 1966 with chance probabilities running from $10^{-4}$
to $10^{-5}$. What is the combined probability of all these previous
cases? And what is the motivation to claim each new case is "a
posteriori"?

\section{The Companion Galaxies Around NGC5985}

\begin{figure}[ht]
\center{\includegraphics[width=.99\hsize]{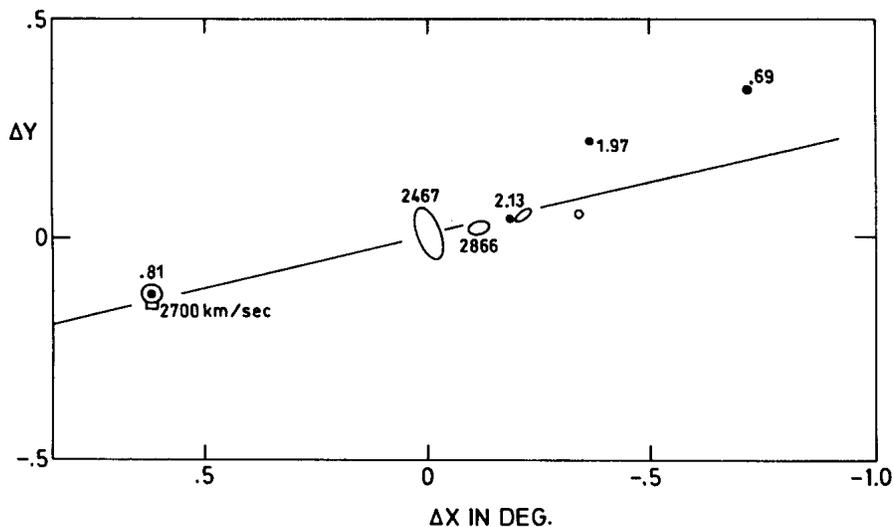}}
\caption{\footnotesize The central portion of Fig. 1 is here enlarged and all 
NGC galaxies are additionally plotted. Three additional companions are seen to 
lie along the WNW minor axis. The redshifts of the galaxies are now labeled 
in km/sec and it is seen that the two companions whose redshifts are known 
have slightly, but significantly, greater redshifts.}
\end{figure}

We have seen that the quasars are aligned along the minor axis of the central 
Seyfert. Sections that follow summarize that generally both quasars and 
companion galaxies are aligned along the minor axes. Is this true of the 
NGC5985 family? In Fig. 2 I have enlarged the central regions of Fig. 1 and 
plotted all the bright, NGC galaxies in the region. It is apparent that these 
NGC galaxies, which are fainter than NGC5985 as companions should be, are 
almost exactly aligned along the WNW extension of the minor axis! Taken 
together with the dwarf companion on the other side of the minor axis, this 
leaves no doubt that in the case of NGC5985 the companion galaxies and quasars
are aligned exactly along the same minor axis. This then constitutes another 
proof that these objects of variously higher redshift have some physical 
relation to the low redshift,central galaxy.

There is confirming evidence in the measured redshifts of these companions. As
Fig. 2 shows, the one to the ESE is about +230 km/sec with respect to NGC5985 
and the one to the WNW is about +400 km/sec. These redshifts are close enough 
to that of the parent galaxy to conventionally confirm their status as physical 
companions. But at the same time they exhibit the systematic excess redshift of 
younger generation galaxies in groups (See Fig. 1 of preceding paper in this 
volume and Arp 1997). In the following sections we will interpret this excess 
redshift as indicating that they have more recently evolved from quasars. Since 
they are older than the quasars, they have had time to fall back from 
apgalacticon to within a few diameters of the parent galaxy.

\section{The Empirical Model of Ejection and Evolution}

As Fig. 3 shows, we can now combine the data from the last 32 years of study of 
physical groups of extragalactic objects. What results is a sequence of quasars 
emerging from a large galaxy along its minor axis, increasing in luminosity and

\begin{figure}[ht]
\center{\includegraphics[width=0.8\hsize]{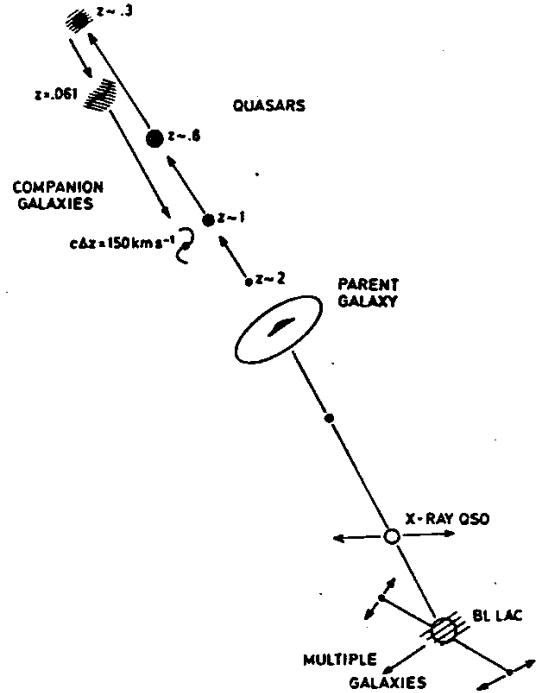}}
\caption{\footnotesize Schematic representation of quasars and companion 
galaxies found associated with central galaxies from 1966 to present. The 
progression of characteristics is empirical but is also required by the 
variable mass theory of Narlikar and Arp (1993).}
\end{figure}

decreasing in redshift as they move outward.  When they reach a maximum distance 
of about 500 kpc they have started to turn into compact, active galaxies. As 
they age further into more normal galaxies they may fall back toward the parent 
galaxy. In that case they fall back along the minor axis because they emerged 
with little or no angular momentum component. {\it They move on plunging orbits.}

Ambartsumian (1958) noted that companion galaxies seemed to be ejected from 
larger galaxies. Arp (1967;1968) showed evidence that quasars were also ejected 
from galaxies. The astronomer most knowledgeable about disk galaxies, Erik 
Holmberg (1969), showed companion galaxies aligned along minor axes. He 
concluded they must be formed from gas ejected from the nucleus. Burbidge and 
Burbidge (1997) showed gas ejected along the minor axis of the strong Seyfert, 
NGC4258. Arp (1997b) showed a number of pairs of X-ray quasars ejected closely 
along the minor axis of Seyferts (e.g. NGC4235, NGC2639).

These observational results are now summarized in Fig. 4. It is seen that the 
quasars preferentially come out in a cone angle of about $\pm20^o$.
 The companion galaxies are preferrentially confined within about $\pm35^o$.
 This difference is in the expected direction because as the quasars reach 
their maximum extension and slow down, they are vulnerable to perturbation by 
objects at that distance and  

\begin{figure}[ht]
\center{\includegraphics[width=0.6\hsize]{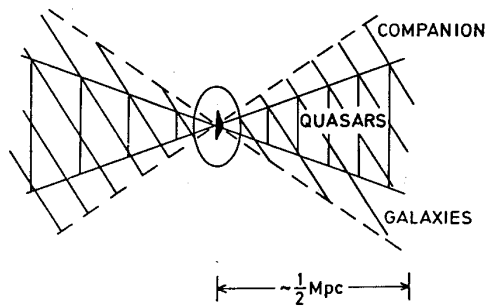}}
\caption{\footnotesize Schematic representation of distribution of companion 
galaxies along minor axes of disk galaxies ($\pm35^o$ from Holmberg 1969; 
Sulentic et al. 1978; Zaritsky et al. 1997). Quasars are observed $\pm20^o$ 
from minor axis (Arp 1998a).}
\end{figure}

hence will fall in again along slightly deviated orbits.  Also since the 
companions are older, the more recent ejection axes of quasars in some cases 
have moved because of precession of the galaxy or the spin axis of the nucleus.

\section{Verification of the Model}

 The above model combines many cases, each one of which may only contain a few 
elements. But it is now possible to show in Fig. 5 the active Seyfert NGC3516. 
This galaxy is exceptional because it confirms the essentials of the model in a 
single case. The objects are drawn from a complete sample of bright X-ray 
objects within about 22 arc min of the Seyfert (Chu et al. 1998). Notice there 
are 5 quasars and a BL Lac-type object, the latter object representing a 
transition between a quasar and a more normal galaxy. The redshifts of these 
six objects are strictly ordered with the highest closest to the Seyfert and 
the lowest furthest away. They define very well the minor axis of the galaxy.

\begin{figure}[ht]
\center{\includegraphics[width=0.8\hsize]{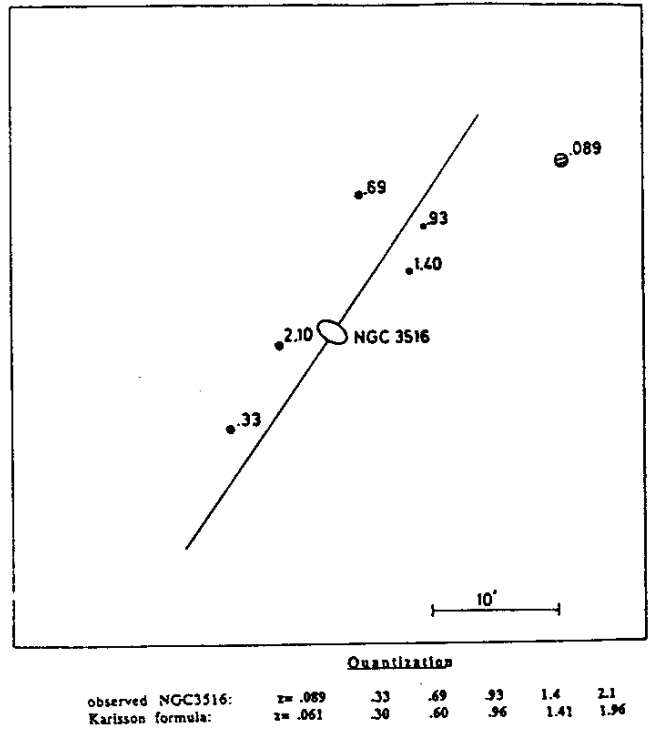}}
\caption{\footnotesize All bright X-ray objects around the very active Seyfert 
galaxy NGC3516 which have had their redshifts measured by Chu et al. (1998). 
Redshifts are written to the upper right of each quasar and quasar like object.}
\end{figure}

 As a side note it is clear that each of the six redshifts fits very closely to 
the quantized redshift peaks observed in quasars in general. Those peaks are:
.06, .30, .60, .96, 1.41, 1.96 (Arp et al. 1990). If these redshifts represented Doppler 
velocities, of course, quantization would be destroyed by random orientations 
to the line of sight. So quantization is direct proof that extragalactic 
redshifts are nonvelocity. The intrinsic redshifts are indicated to evolve in 
discrete steps as the quasars evolve into galaxies. The quantization of 
redshifts in smaller steps in the companion galaxies further supports their 
being end products of this evolution.

The NGC5985 case shown in Fig. 2 confirms very well the later stages of smaller 
redshift where the evolved companion galaxies have fallen back toward the 
parent galaxy. The very exact alignment of quasars and companion galaxies must 
mean that the minor axis of NGC5985 has stayed unusually well fixed in space 
over the time required to evolve from a high redshift quasar to a reasonably 
normal galaxy, or about 7 x 109 years (Arp 1991). Another case of a very narrow 
alignment of companion galaxies can be seen around our Local Group center, M31 
(Arp 1998a).

The length of the NGC5985 line and the brightness of its components suggests 
that the group is closer than most Seyferts so far investigated. In particular 
the distribution of quasars is much closer to NGC3516 which appears to be more 
distant. It is also clear, however, from the measured ellipticities of the 
images, that the minor axis of NGC3516 is tipped much closer to our line of 
sight than in NGC5985. That accounts for some foreshortening of the NGC3516 
line as well as an apparently greater spread off the line relative to the 
length of the line.

\section{Statistics} 

 From 1966 onward statistical associations of high redshift quasars with low 
redshift galaxies were reported (Arp 1987). Significances ranged from 16 to 
greater than 7.5 sigma. We will give only the latest example here as shown in 
Fig. 6. There it is shown that a nearly complete sample of Seyfert galaxies 
have a conspicuous excess of bright X-ray point sources within about 50 arcmin 
radius (Radecke 1997; Arp 1997a). Since these X-ray sources are essentially all 
quasar or quasar-like one can see at a glance that there are a large number of 
quasars physically associated with nearby Seyfert galaxies.

\begin{figure}[ht]
\center{\includegraphics[width=1.0\hsize]{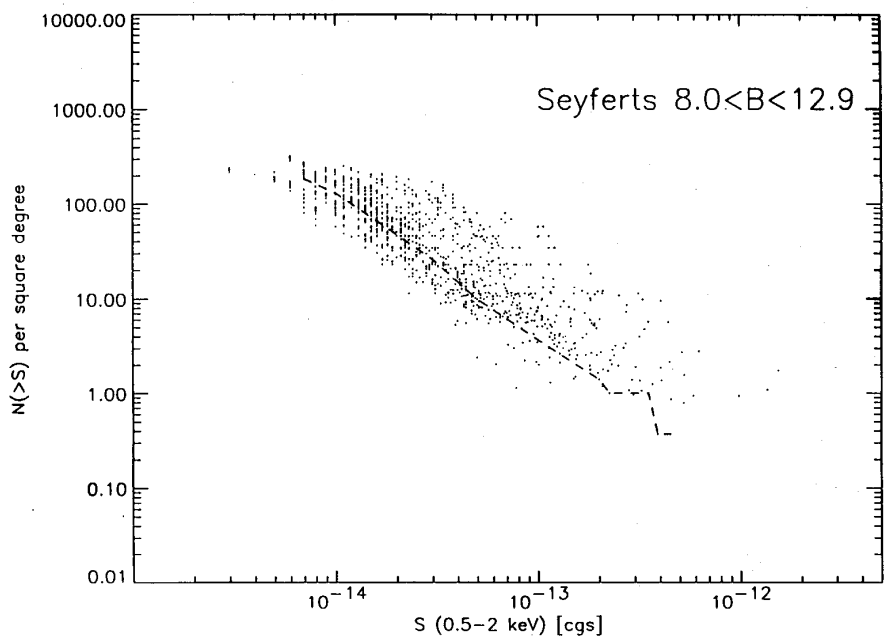}}
\caption{\footnotesize Cumulative number of X-ray sources brighter than 
strength (S) around a nearly complete sample of bright Seyfert galaxies. 
Dashed line represents counts in non-galaxy control fields.}
\end{figure}

\section{Theory} 
 What would give an intrinsic (non-velocity) redshift to a galaxy - high when 
it is first born - and decreasing as it ages? The answer was supplied by 
Narlikar in 1977. It rests on the fact that the standard solution of the 
Einstein field equations as made by Friedmann in 1922 made one key assumption. 
It assumed that the elementary particles which constitute matter remain forever 
constant in time. A more general solution, as Narlikar showed, had the mass of 
particles increasing with time. This would furnish a series of answers to 
paradoxes which currently falsify the standard Big Bang solution: 
     
     1) Episodic creation near zero mass, giving initial outflow at signal 
        speed, c.  
   
     2) Condensation into proto galaxies. 
     
     3) An intrinsic redshift initially high and decreasing with time. 

     4) Evolution into normal galaxies moving with small velocities. 

     5) Possibility of understanding quantized redshifts. 
     
     6) Transformation to local physics using local time scales.

The details of this theory are explained in recent publications Arp (1998b). 
But here it should be emphasized that the observational disproofs of the 
conventional, singular creation theory can no longer be discarded with the 
claim that there is no viable alternative theory. The empirical conclusions of 
Ambartsumian and many others who followed, have now been supported by numerous 
amplifying observations and a unifying physical theory has been adduced.

It is clear from the experience of the last 40 years that influential 
astronomers will accept neither the observations which falsify the current 
beliefs nor the theory which enables these observations to be understood. 
Therefore it is of the utmost importance for each individual researcher to 
examine and judge the facts for themselves. The usefulness of each person's 
career labors and their usefulness to science will depend on their choosing the 
correct fundamental assumptions.

\vskip3ex
\centerline{References}
\noindent{\footnotesize

Ambartsumian, V.A. 1958, Onzieme Conseil de Physique Solvay, ed. R. Stoops, 
Bruxelles. 

Arp, H. 1967 Ap.J. 148, 321. 

Arp, H. 1968, Astrofizika 4, 59. 

Arp, H. 1987, "Quasars, Redshifts and Controversies" Interstellar Media, 
Berkeley 

Arp, H. 1991, Apeiron Nos. 9-10, 18. 

Arp, H. 1997, Astrophysics and Space Sciences 250, 163. 

Arp, H. 1997a, Astron. Astrophys. 319, 33. 

Arp, H. 1997b, Astron. Astrophys. 328, L17. 

Arp, H. 1998a, Ap.J. 496, 661. 

Arp, H. 1998b, "Seeing Red: Redshift, Cosmology and Acad. Sci." Apeiron, 
Montreal 

Arp, H., Bi, H., Chu Y., Zhu, X. 1990, Astron. Astrophys. 239, 33.

Burbidge, G.R., Burbidge, E.M. 1997, Ap.J. 477, L13. 

Chu, Y., Wei, J. Hu, J., Zhu, X. and Arp, H. 1998, Ap.J. 500, 596. 

Holmberg, E. 1969, Arkiv of Astron., Band 5, 305. 

Markarian B.E., Stepanyan D.A., Erastova L.K. 1986 Astrophysics 25, 51 

Narlikar, J.V. and Arp, H. 1993, Ap.J. 405, 51. 

Radecke, H.-D. 1997, Astron. Astrophys. 319, 18. 

Reimers D., Hagen H.-J. 1998, Astron. Astrophys. 329, L25. 

Sulentic, J.W., Arp, H. and Di Tullio, G.A. 1978, Ap.J. 220, 47.

Zaritsky, D., Smith, R., Frenk, C.S. and White, S.D.M. 1997, Ap.J. 478, L53.}

\end{document}